\documentclass[a4paper,fleqn,usenatbib]{mnras}

\usepackage{newtxtext,newtxmath}
\usepackage{color}

\usepackage[T1]{fontenc}
\usepackage{ae,aecompl}

\usepackage{graphicx}	
\usepackage{url}
\usepackage{amsmath}	
\usepackage{subfigure}
\usepackage{makecell}

\title[TELNET]{Point Spread Function Estimation for Wide Field Small Aperture Telescopes with Deep Neural Networks and Calibration Data}

\author[Peng Jia et al.]{
Peng Jia$^{1, 4}$\thanks{robinmartin20@gmail.com},
Xuebo Wu$^{1}$, Zhengyang Li $^2$ \thanks{zyli@niaot.ac.cn}, Bo Li$^{2}$, Weihua Wang$^{1}$, Qiang Liu$^{1}$, \\
\and Adam Popowicz$^3$ \thanks{adam.popowicz@polsl.pl} and Dongmei Cai$^{1}$\\
$^{1}$College of Physics and Optoelectronics, Taiyuan University of Technology, Taiyuan, 030024, China\\
$^{2}$Nanjing Institute of Astronomical Optics and Technology CAS, Nanjing, Jiangsu, 210042, China\\
$^{3}$Department of Electronics, Electrical Engineering and Microelectronics, Silesian University of Technology,  Akademicka 16, 44-100 Gliwice, Poland\\
$^{4}$Department of Physics, Durham University, DH1 3LE, UK\\
}

\date{Accepted XXX. Received YYY; in original form ZZZ}

\pubyear{2021}

\begin{document}
\label{firstpage}
\pagerange{\pageref{firstpage}--\pageref{lastpage}}
\maketitle

\begin{abstract}
The point spread function (PSF) reflects states of a telescope and plays an important role in development of data processing methods, such as PSF based astrometry, photometry and image restoration. However, for wide field small aperture telescopes (WFSATs), estimating PSF in any position of the whole field of view is hard, because aberrations induced by the optical system are quite complex and the signal to noise ratio of star images is often too low for PSF estimation. In this paper, we further develop our deep neural network (DNN) based PSF modelling method and show its applications in PSF estimation. During the telescope alignment and testing stage, our method collects system calibration data through modification of optical elements within engineering tolerances (tilting and decentering). Then we use these data to train a DNN (Tel--Net). After training, the Tel--Net can estimate PSF in any field of view from several discretely sampled star images. We use both simulated and experimental data to test performance of our method. The results show that the Tel--Net can successfully reconstruct PSFs of WFSATs of any states and in any positions of the FoV. Its results are significantly more precise than results obtained by the compared classic method - Inverse Distance Weight (IDW) interpolation. Our method provides foundations for developing of deep neural network based data processing methods for WFSATs, which require strong prior information of PSFs.
\end{abstract}




\section{Introduction}
Telescopes with wide FoV (normally with a diameter greater than 1 deg) and small apertures (less than 2 metre) are called WFSATs. They are low-cost, light-weighted and commonly used for wide field sky surveys in time domain astronomy \citep{burd2005pi, Ma2007, yuan2008chinese, cui2008antarctic, Pablo2016,ping2017the,ratzloff2019building,sun2019precise}. Because apertures of WFSATs are small, they are normally used to observe bright targets. To further increase detection and observation abilities of WFSATs, new data processing methods with strong prior conditions have recently been developed and proven their superiority over existing image processing techniques \citep{Jia2020, Glazier2020}.\\ 

The PSF is the impulse response of an optical system and it reflects states of the whole telescope.  For optical observation data processing methods, PSF is a commonly used prior condition. Prior conditions of PSFs mainly include shape and variations of PSFs within the whole FoV. For general purpose sky survey telescopes, different data processing methods have been recently tested on images with various PSFs and have successfully proved its robustness \citep{Gonz2018, Duev2019, Burke2019, He2020}. However, PSFs of WFSATs are quite different from that of general purpose sky survey telescopes, because WFSATs are often working remotely (lack of maintenance) and images obtained by them have both low sampling rate and fluctuating quality (e.g. due to the high sensitivity of some optical elements to temperature and/or humidity). PSFs of WFSATs have significant deformation in different regions of FoV \citep{Piotrowski2010} and they can evolve with time \citep{Jia2017}. It is very hard to design a fully deterministic image processing method that can model spatially and temporally PSFs, so more sophisticated PSF estimation methods are required.\\

The statistical PSF modelling method is commonly used for WFSATs nowadays. The statistical methods require large amount of star images as samples of PSFs and obtain effective components from these images through either principal component analysis \citep{Jee2007, Jia2017, Popowicz2018, sun2020improving} or denoising autoencoders \citep{Jia2020b}. Because large amount of star images with adequate signal to noise ratio (SNR) are required \citep{Wang2018}, application of such methods is frequently unsuitable for WFSATs that are used for fast sky survey.\\

The forward PSF modelling method, which was firstly applied in space based telescopes \citep{Krist2011,Perrin2012} and later in telescopes with adaptive optics systems \citep{martin2016psf, fetick2019physics, beltramo2019prime, fusco2020reconstruction,  Jia2021}, shows good performance in modelling PSFs generated by the atmospheric turbulence with the help of DNN \citep{ Jia2020SPIE,jia2020non}. The forward DNN based PSF modelling method (PSF--NET) builds the Monte Carlo model of the imaging process and generates huge amount of PSFs to train the DNN. After training, the PSF--NET learns a function that can reflect the repose of a telescope (PSF) to complex external environments (such as the shape of wavefront distortion due to the atmospheric turbulence).\\

For WFSATs, because all lenses and mirrors are manufactured and assembled with predefined tolerances, PSFs of WFSATs can be interpreted as the response of an imaging system to the actual state of optical misalignment together with the actual quality of fabricated optical elements  \citep{li2015alignment}. The distribution of PSFs with different shapes also reflects the impact of external conditions during imaging, like temperature, humidity or wind strength, which can lead to very complex deformations of PSFs \citep{Jia2020c}. Therefore, we decided to design the PSF estimation technique using current capabilities of artificial intelligence since such algorithms have already proven to be successful in realization of complex tasks in plenty of various fields.\\

In this paper, we propose the Tel--Net as a novel PSF modelling method for WFSATs. With tolerances of each optical element in a WFSAT as prior information, the Tel-Net is trained on various possible realizations of the optical alignment. After training, the Tel--Net can output PSFs of a WFSAT in any state and in any position within the FoV. We introduce the concept and structure of the Tel--Net in Section \ref{sec:telnet}. The performance of the Tel--Net is tested with simulated data in Section \ref{sec:simresult} and with real data in Section \ref{sec:realresult}. We give the conclusions and propose the future work in Section \ref{sec:con}.\\

\section{The Tel-Net}
\label{sec:telnet}
\begin{figure*}
   \centering
   \includegraphics[width=\hsize]{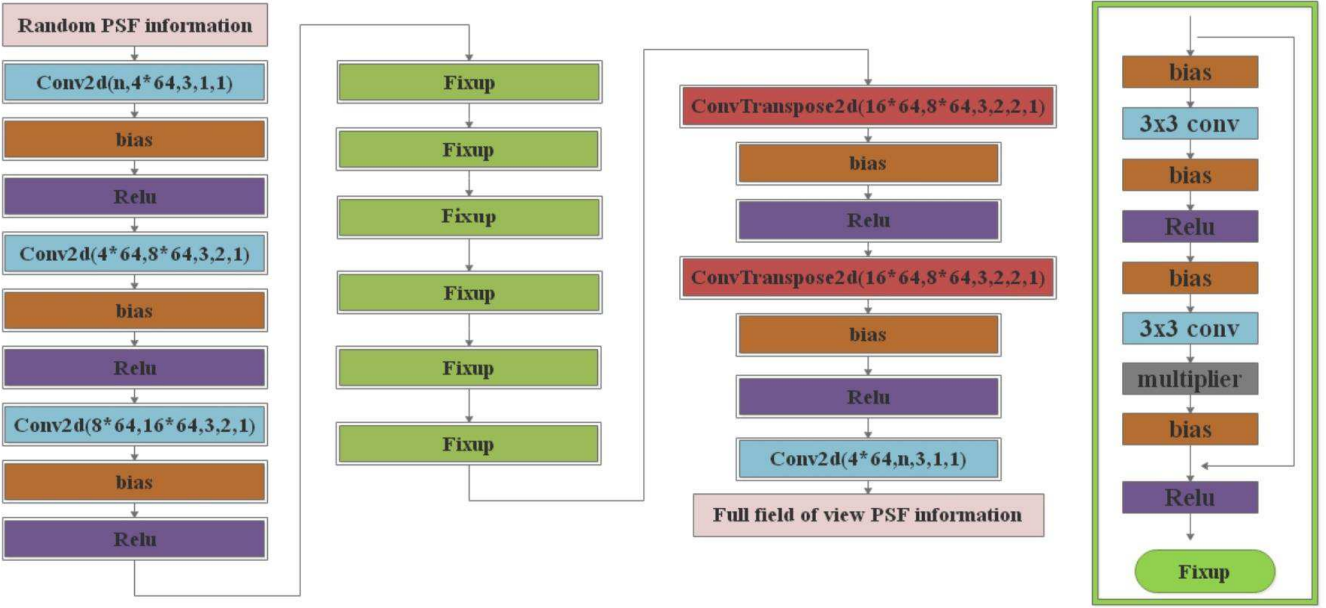}
      \caption{The structure of the Tel--Net. The Tel--Net consists of convolutional layers (Conv2d in blue), Bias layers (Bias in brown), Activation layers (Relu in purple), Conv-transpose layers (ConvTranspose2d in red), multiply layers (multiplier in grey) and Fixup blocks (Fixup block in green). The structure of Fixup block is shown in the right part of this figure.}
   \label{figure001}
   \end{figure*}

The philosophy of the Tel--Net is quite simple. PSFs in the whole FoV are representations of the telescope state \citep{li2015alignment} which can be modelled by a DNN \citep{Jia2020c}. After we train the Tel--Net with some PSFs as inputs and all PSFs within the whole FoV as outputs, the Tel--Net will be able to output all PSFs of a WFSAT in different states within the whole FoV with only some PSFs given at its input. The structure of the Tel--Net is adopted from the PSF--NET proposed in \citet{Jia2020c} and the DAE--NET proposed in \citet{Jia2020b}. As shown in figure \ref{figure001}, the Tel--Net has an encoding and decoding structure with DNN blocks adopted to increase its representation ability.\\

Basic components in the encoding part of Tel--Net is convolutional neural network (CNN). The CNN uses trainable shared-weight architecture to extract features from two multi-dimensional data, which makes it suitable for multi-dimensional data processing, particularly for images. The CNN is widely used for tasks such as: astronomical target detection \citep{gonzalez2018galaxy, burke2019deblending, Jia2020, hausen2020morpheus}, classification \citep{jia2019optical, turpin2020vetting, He2020} and image restoration \citep{schawinski2017generative,jia2019solar, jia2021data}. In the Tel--Net, the CNN is used to extract features of PSFs between pixels in the same PSF and pixels of different PSFs in different positions of FoV. A bias layer and a Relu layer are added after each convolutional layer. The bias layer adds additional values to outputs of the previous layer to make the activation function of the Relu layer effective. The Relu layer would applies function $f(x)=max(0,x)$ to inputs and it would add non--linear properties to the DNN (where $x$ stands for the input of Relu layer and $max$ stands for function to return maximal value between $x$ and 0).\\

In the decoding part of the Tel--Net, transposed convolutional layer ($ConvTranspose2d$), bias layer and Relu layer are arranged to form basic structure. The transpose convolutional layer also has shared-weight architecture and it generates outputs with dimensions that are larger than dimensions of inputs. With the encoding and decoding structure, the Tel--Net can estimate PSFs in different positions of the FoV from input PSFs.\\

The deeper the neural networks are, the better it would be able to model complex functions \citep{goodfellow2016deep}. To increase the representation ability of the Tel--Net, we introduce Fixup blocks in the Tel--Net as DNN blocks. The structure of Fixup blocks is shown in right panel of figure \ref{figure001}, which uses fixed--update initialization, also known as Fixup method proposed by \citet{zhang2019fixup}. The Fixup method is a regularization method that rescales weights of all layers in DNNs properly at the beginning of training. Comparing with other ordinary regularization methods, such as the normalization layers \citep{ba2016layer}, the Fixup method can accelerate convergence and improve generalization without changing greyscale values of input images. Introduction of the Fixup method would be benefit in applications of astronomical image processing, because grey values of astronomical images have a wider range than that in typical 8-bit images and original grey values instead of their distributions are important for scientific data processing. \\

For the Tel--Net, PSFs of WFSATs in a given state are represented by a data cube (PSF-Cube) with dimensions of $N\times N\times M$, where $N$ stands for width and length of star images and $M$ stands for the number of PSFs that are used to represent a given FoV. In this paper, we divide the FoV of WFSATs into $M$ small sections with equal size (called patches) and we assume that PSFs are uniform in each patch \citep{la2015method}. These PSFs are arranged from left to right and top to bottom according to positions of patches in the FoV. Each PSF in a PSF-Cube is aligned with its brightest pixel in the centre. Besides, to reduce effects brought by sub-pixel uncertainties and to extend the training set, we randomly shift PSFs by a fraction of a pixel (within -0.5 to 0.5 pixels with uniform probability distribution) when we obtain PSF--Cubes.\\

The input and output of the Tel--Net are both PSF-Cubes of the same size. However, there are only several PSFs in input PSF-Cubes and others are zero matrices. For output PSF--Cubes, all matrices are PSFs. Because the Tel--Net should be able to predict all PSFs from several known PSFs, the input and output PSF--Cubes that correspond to the same state of the WFSAT are used as a pair of training data for the Tel--Net. During the training stage, the batch size is set to 4 (due to limited available memory in the GPU -- GTX 1080 in this paper) and we choose the Adam Optimizer \citep{kingma2014adam} as the optimization algorithm. The learning rate is set as 0.01 and the loss function -- $L_1$ norm is defined in equation \ref{eq:equation1}, where $n$ equals to the batch size, $TelNet$ $(PSFCube_{in})$ is the output PSF-Cube obtained from the Tel--Net and the $PSFCube_{out}$ is expected output PSF--Cube, the norm $||$ utilizes the absolute difference, $n$ stands for the total number of PSFs in the PSF-Cube and $k$ stands for index of these PSFs.\\
\begin{equation}\label{eq:equation1}
L=\frac{1}{n}\sum_{k=1}^n |TelNet(PSFCube_{in}) - PSFCube_{out} | \\
\end{equation}

There are two possible ways to obtain PSFs to train the Tel--Net: computation--based PSFs or data--driven PSFs. We can obtain PSF of any position in the whole FoV for a WFSAT in any state through physical computation \citep{Jia2020c}. However, because there are differences between PSFs obtained by physical computation and PSFs in real observed images, the former can only be used as pre--trained data. PSFs that are obtained from real observations (data--driven PSFs) are required for real applications. Due to hardware limitations or time consumptions in obtaining PSFs in any position of the FoV during real observation, data--driven PSFs can only be obtained for a WFSAT with finite states in several predefined positions in the FoV and with finite accuracy. These problems will be further discussed in Section \ref{sec:simresult} and Section \ref{sec:realresult}.\\

\section{Training and testing the Tel-Net with simulated data}
\label{sec:simresult}
\begin{figure*}
   \centering
   \includegraphics[width=0.7\hsize]{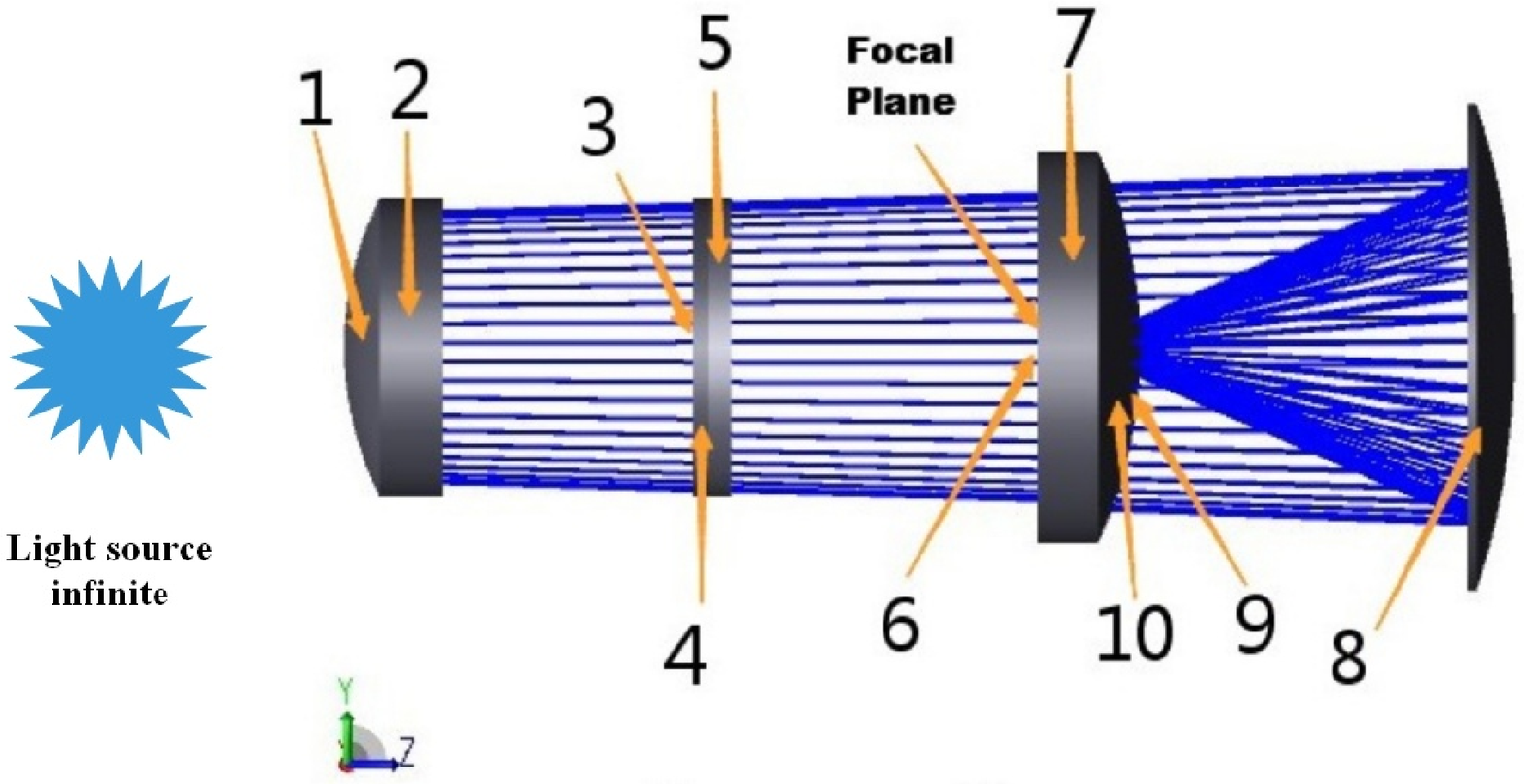}
	\begin{tabular}{cccccc}
		\hline
		Surface & Surface Type &  Radius (mm) & Separation (mm) & Glass & Semi-Diameter (mm)\\
		\hline
		1 & Spherical & 225.0 & 42.8 & BK7 & 164.5 \\
		2 & Spherical & 191.3 & 197.5 & Air &143.8\\
		3 & Even Aspherical & Infinity & 14.0 & F2 & 98.6 \\
		   & (0, $-5.6  \times 10^{-10}$, $-2.4 \times 10^{-14}$) & & &  &\\
	      4 &  Spherical & -1355.0 & 13.0 & SK2 & 101.1 \\
	      5 & Plane & Infinity & 239.2 & Air &105.1 \\
	      6 & Spherical & -293.1 & 42.8 & BK7 & 177.5 \\
	      7 & Spherical & -324.2  & 251.0 & Air & 196.3 \\
	      8 & Spherical & -550.0 & -251.0 & Mirror & 290.8 \\
	      9 & Spherical & -324.2 & -42.8 & BK7 & 119.9\\
	      10 & Spherical & -293.1 & -14.5 & Air & 95.1\\
	      Focal Plane & Spherical & -241.8 & - & - & 82.6\\
		\hline 						
	\end{tabular}	
      \caption{Parameters of a simulated WFSAT with Baker super-Schmidt design. The third surface is an even aspherical surface and the aspherical coefficients are also shown in this table}
   	\label{figure002}
   \end{figure*}

In this section, we simulate a classic WFSAT (Baker super-Schmidt telescope) with ZEMAX to test the performance of the Tel--Net. Parameters of this WFSAT are shown in figure \ref{figure002}. This WFSAT has a field of view of $10^{\circ}$ and we divide it equally into $61\times 61$ patches. In each patch, variations of PSFs are small thus a single PSF is representative. There are four optical elements in this telescope and we assume that the second lens has alignment error. There are six free dimensions for the second lens: decenter and tilt along x, y and z directions. To make sure that star images in different places within the FoV can be obtained by the WFSAT, we set the tolerance of the second lens as: -3 millimetres to 3 millimetres with step of 1.5 millimetres for x, y and z decenter, -0.2 degree to 0.2 degree with step of 0.1 degree for x and y tilt. In total, there are 2000 different states for this element and we could obtain 2000 PSF-Cubes. In each PSF-Cube, there are $61 \times 61$ PSFs and we obtain 744200 PSFs in total as the training set of the Tel--Net. The size of PSF window is 15.33 um with $64\times 64$ pixels and pixelscale of 0.27 arcsec/pixel. Meanwhile, we generate 100 PSF-Cubes with the same WFSAT and this time we set the second mirror in random states within its misalignment configurations as the test set.\\

It should be noted that in essence PSF-Cubes in the training set are sparse samples of all PSFs that correspond to the WFSAT within tolerances of its secondary mirror. Because we sampled tolerances with relatively big steps, there are still large number of PSFs that can be obtained. Assuming all elements have the same number of free states as the second element, there would be tens of thousands of PSF--Cubes to be obtained, which is impossible to be processed in real applications. However, for PSF estimation task, many of these states are redundant (although the telescope belongs to two different states, PSFs in a given point within FoV are almost the same). This is why we decided to make a trade--off between complexity and accuracy, and so we use sparsely sampled PSF--Cubes to train the Tel--Net. In the future, it would be necessary to further investigate possible methods of the optimal reduction (minimization) of the training set to its most informative subset.\\

\subsection{Training and testing the Tel-Net with simulated data of small size}
\label{sec:simresultsmall}
First of all, we test the robustness of the Tel--Net using a small data set, which includes only 200 PSF--Cubes reflecting states of the WFSAT along tilt of x and y and decenter of x. We select 150 from 200 PSF--Cubes to train the Tel--Net. For the input PSF--Cubes, we randomly select 60 PSFs as known PSFs and set all other PSFs to zero matrices. For the output PSF--Cubes, we use the same PSF--Cube with all PSFs as valid PSFs. After training, we select the remaining 50 PSF-Cubes to test the Tel--Net.\\

Meanwhile we use direct interpolation method for comparison. We select the Inverse Distance Weight (IDW) interpolation method \citep{lu2008adaptive} as exemplary classic interpolation method. The IDW interpolation method is a commonly used weighted average interpolation method in which the weights depend on the reciprocal of the distance between interpolated points. We use the mean square error (MSE) between predicted PSFs and original PSFs as the evaluation function.\\

\begin{figure}
   \centering
   \includegraphics[width=\hsize]{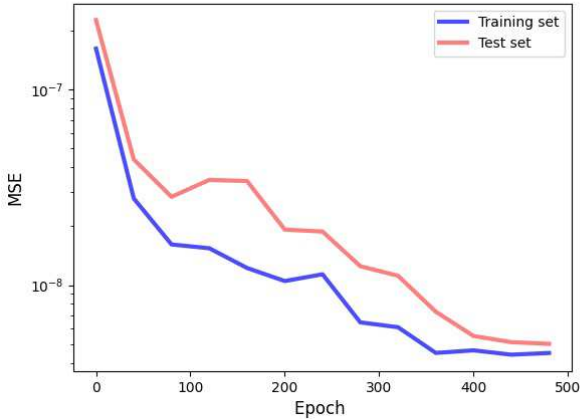}
      \caption{MSE of PSFs in the training set and the test set as function of epochs, when the training set has smaller number of PSFs.}
   \label{figure003}
   \end{figure}

Since the Tel--Net is very deep and the number of PSF-Cubes is small, there would be some risks that the Tel--Net would over--fit these PSFs. Therefore, we plot the MSE of PSFs (both in the training set and the test set) as function of epochs as shown in figure \ref{figure003}. In order to save storage space, we save weights for every 10 epochs. We have trained the Tel--Net for 480 epochs in total and calculate the mean MSE for every 4 epochs. From this figure, we can conclude that the Tel--Net could obtain effective representation of PSFs without over--fitting.\\
	\begin{table}
	\caption{MSE between predicted PSFs (estimated by the Tel--Net and the IDW) and original PSFs. The Tel--Net is trained by 150 training PSF--Cubes and tested by other 50 test PSF--Cubes.}                 
	\centering          
	\begin{tabular}{c c c c}     
	\hline\hline                         
	Method &MSE mean&MSE variance \\
	\hline   
	Tel-Net &$3.77  \times 10^{-8}$&$2.76  \times 10^{-16}$\\
	IDW &$8.01  \times 10^{-7}$&$2.09  \times 10^{-13}$\\
	\hline           
	\end{tabular}
	\label{table:PSFMSE1}
	\end{table}

\begin{figure}
   \centering
   \includegraphics[width=\hsize]{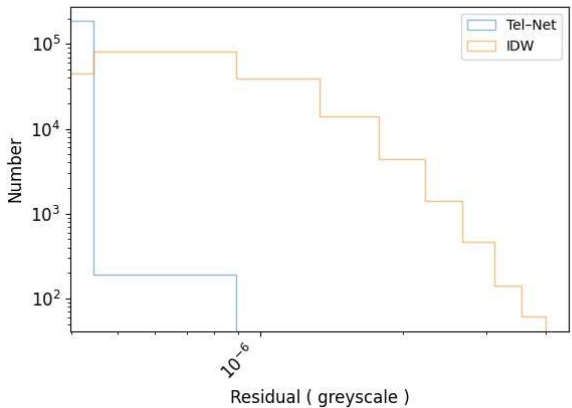}
      \caption{Histogram of MSE between PSFs predicted by Tel--Net and original PSFs. The blue histogram stands for MSE of PSFs predicted by the Tel--Net and orange histogram stands for MSE of PSFs predicted by the IDW method. The Tel--Net is trained by 150 training PSF--Cubes and tested by the other 50 test PSF--Cubes.}
   \label{figure004}
   \end{figure}

\begin{figure*}
\centering
\subfigure[The heat map of mean MSE of PSFs reconstructed by the Tel--Net]{
\begin{minipage}[t]{0.45\linewidth}
\centering
\includegraphics[width=1\textwidth]{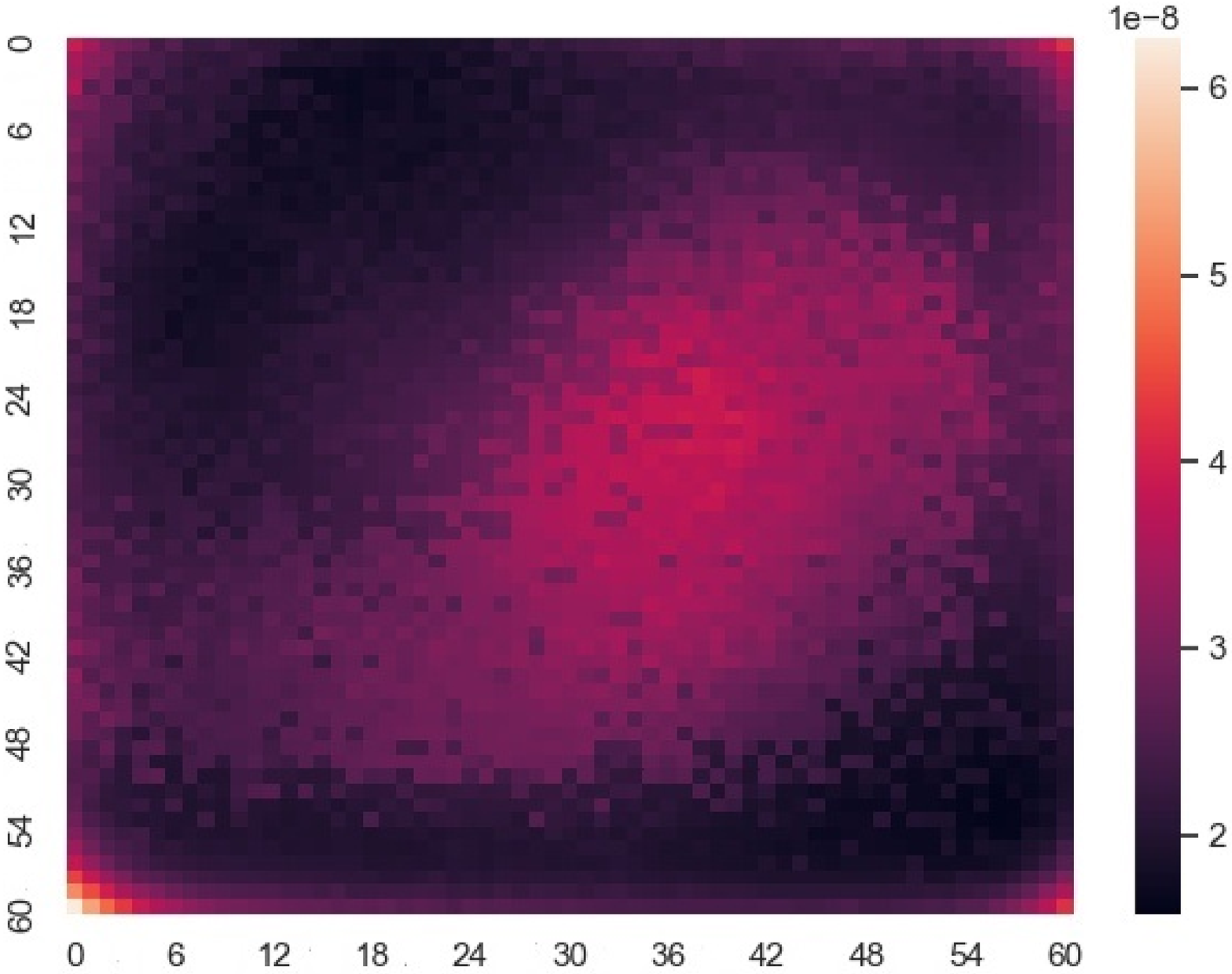}
\end{minipage}%
}%
\subfigure[The heat map of mean MSE of PSFs reconstructed by the IDW]{
\begin{minipage}[t]{0.45\linewidth}
\centering
\includegraphics[width=1\textwidth]{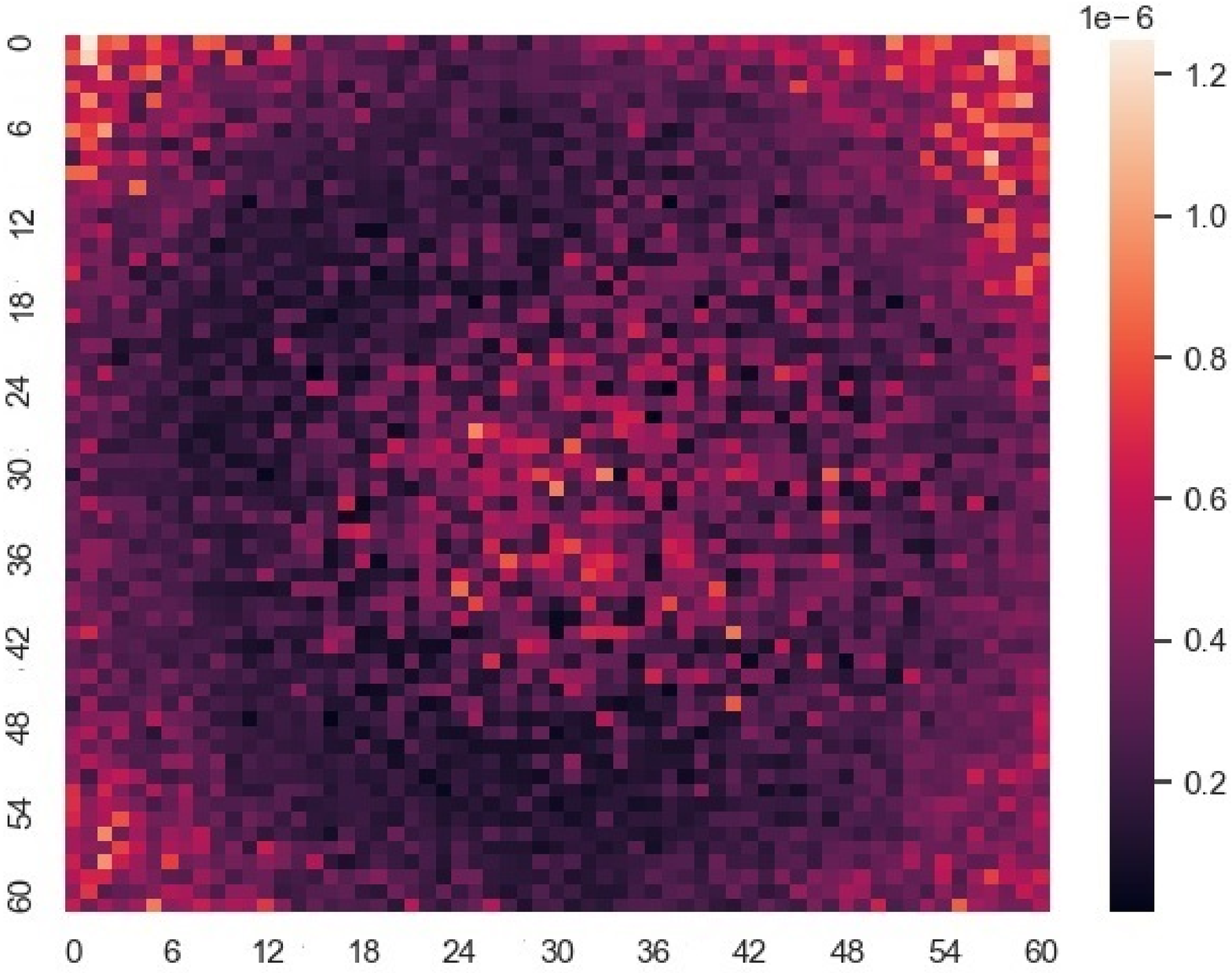}
\end{minipage}%
}%
\centering
\caption{The heat map of mean MSE between original PSFs and PSFs reconstructed by different methods.}
\label{figure005}
\end{figure*}

\begin{figure*}
\centering
\subfigure[Original PSF]{
\begin{minipage}[t]{0.33\linewidth}
\centering
\includegraphics[width=1\textwidth]{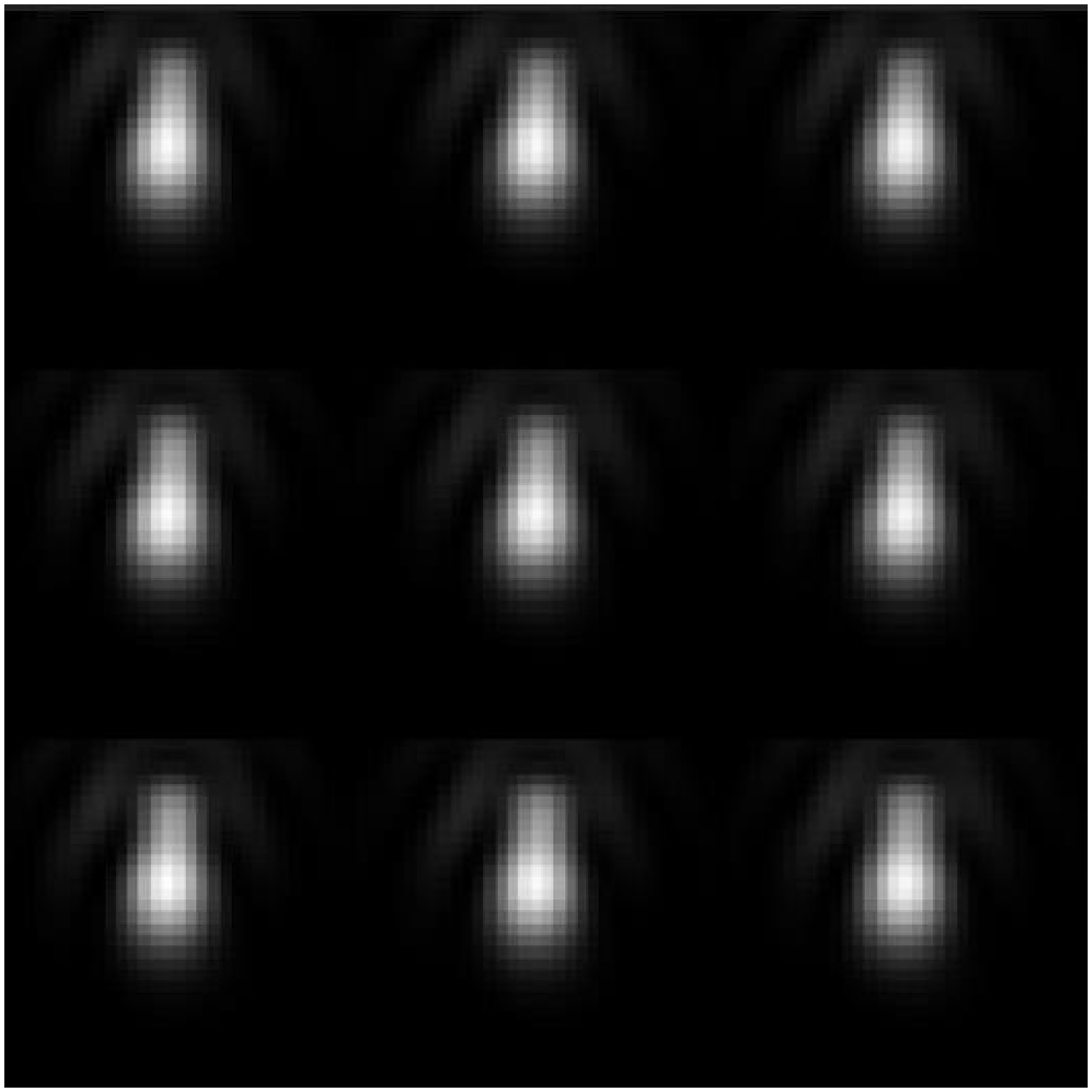}
\end{minipage}%
}%
\subfigure[Tel-Net predicted PSF]{
\begin{minipage}[t]{0.33\linewidth}
\centering
\includegraphics[width=1\textwidth]{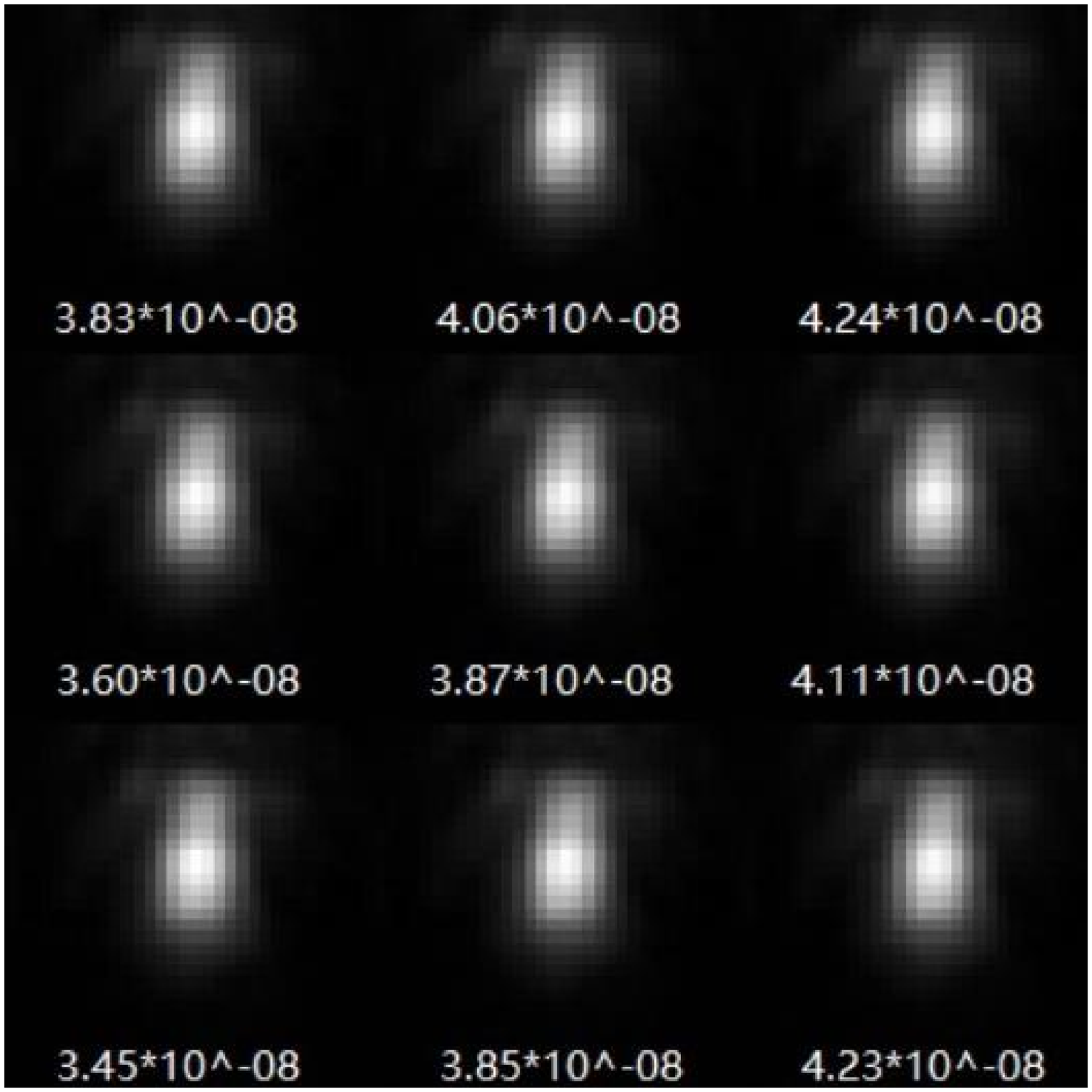}
\end{minipage}%
}%
\subfigure[IDW predicted PSF]{
\begin{minipage}[t]{0.33\linewidth}
\centering
\includegraphics[width=1\textwidth]{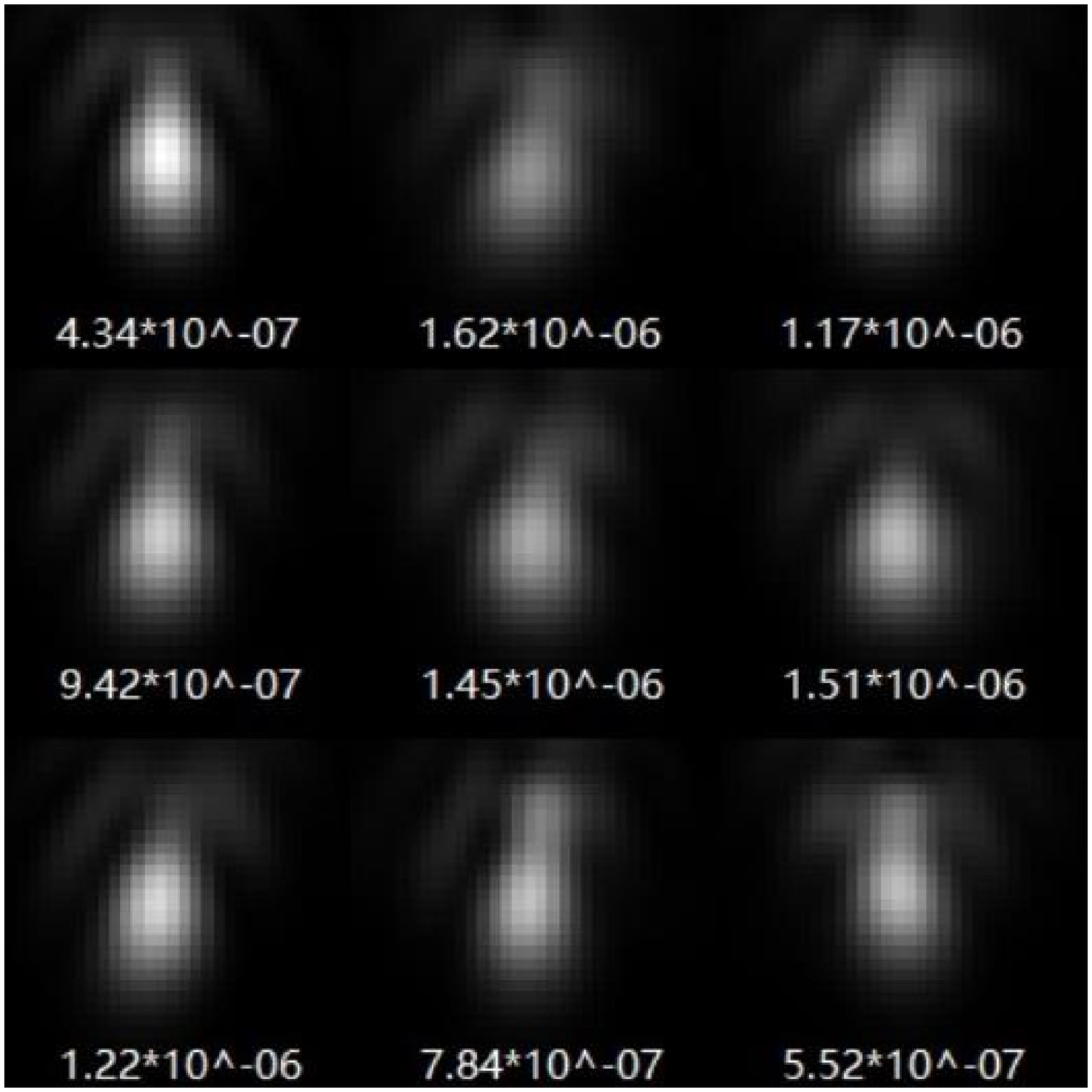}
\end{minipage}
}%
\centering
\caption{Several examples of original PSFs and corresponding PSFs reconstructed by different methods with their MSE values, when the number of PSFs in the training set is small. (a) are original PSFs, (b) are PSFs predicted by the Tel--Net and (c) are PSFs predicted by the IDW with with four nearby PSFs.}
\label{figure006}
\end{figure*}

We use the trained Tel--Net to reconstruct PSFs for WFSATs. The statical results of the MSE are shown in table \ref{table:PSFMSE1} and histograms of the MSE are exposed in figure \ref{figure004}. We can find that with only a limited number of PSFs, the Tel--Net could still be able to predict PSFs with relatively high accuracy, while the existing IDW method presents a magnitude higher MSE. We further plot the heat map of MSE between PSFs reconstructed by different methods and original PSFs in the whole FoV as shown in figure \ref{figure005}. We can find that the MSE of reconstructed PSFs is smaller in the centre of the FoV and PSFs in corner of the FoV are not well reconstructed. This trend is the same for the IDW method, because in the corner there would be less reference PSFs. But the MSE of PSFs reconstructed by the Tel--Net is smaller than PSFs reconstructed by IDW method. One of original PSFs and PSFs reconstructed by different methods are shown in figure \ref{figure006}. We can conclude the following results:\\
\begin{itemize}
  \item [1)] 
 PSFs of WFSATs tested in this part have irregular shapes;
  \item [2)]
  PSFs obtained by the IDW method can not retain complex shapes, which is probably the reason why our method has much better prediction results than that of the IDW method;
  \item [3)]
  Tel-Net has both low mean MSE and low variance of MSE which indicates high repeatability of good results of this estimator.
\end{itemize}

\subsection{Training and testing the Tel-Net with simulated data of big size}
\label{sec:simresultbig}
\begin{figure}
   \centering
   \includegraphics[width=\hsize]{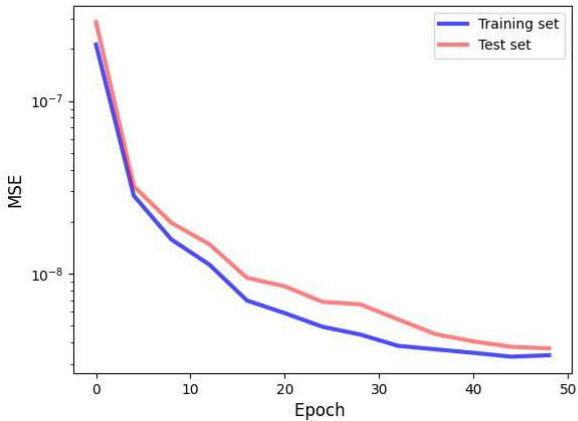}
      \caption{MSE of PSFs in the training set and the test set as function of epochs, when the training set has bigger number of PSFs.}
   \label{figure007}
   \end{figure}

\begin{figure}
   \centering
   \includegraphics[width=\hsize]{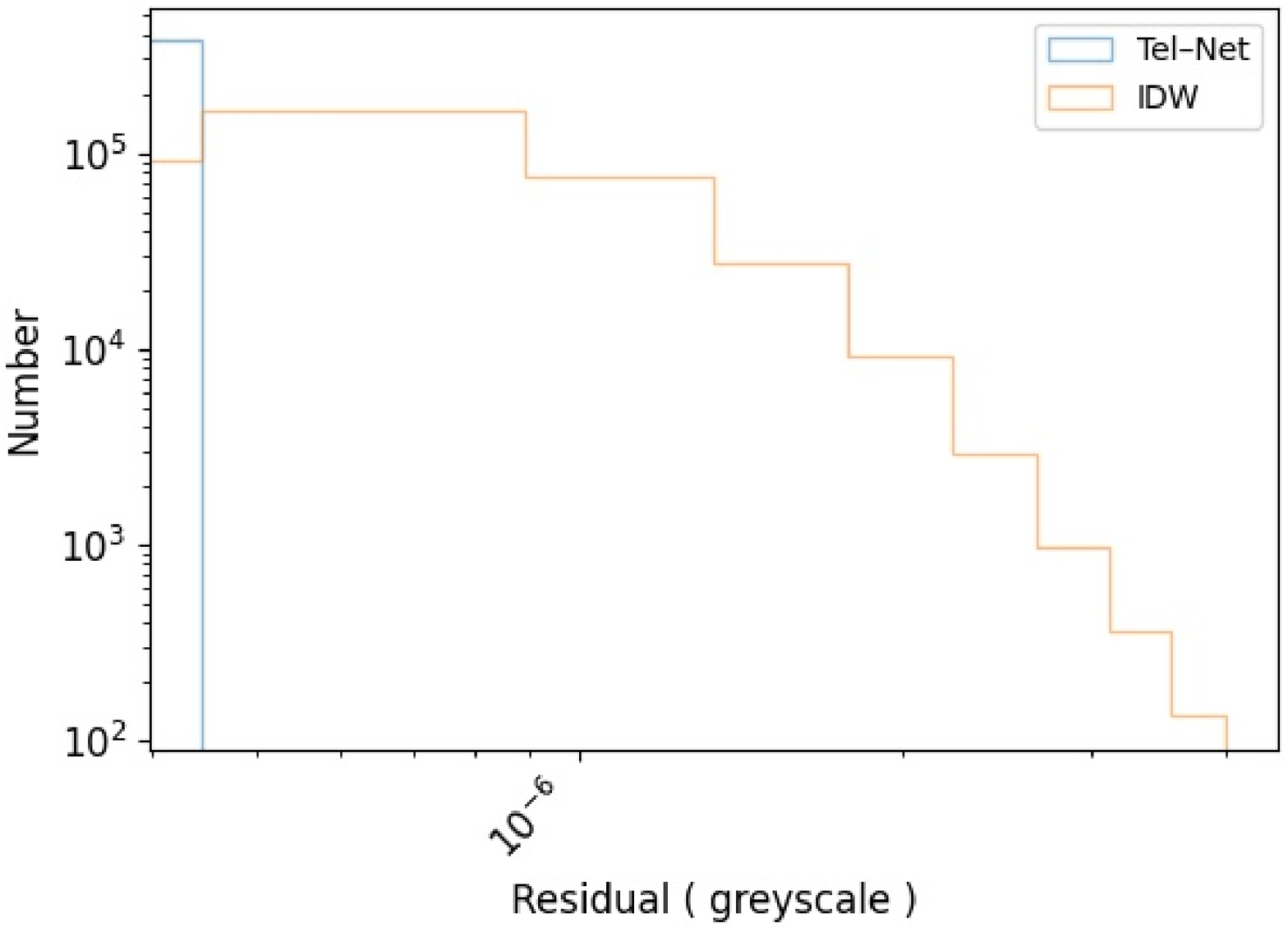}
      \caption{Histogram of MSE between PSFs predicted by Tel--Net and original PSFs. The blue histogram stands for MSE of PSFs predicted by the Tel--Net and orange histogram stands for MSE of PSFs predicted by the IDW method. The Tel--Net is trained by 2000 training PSF--Cubes and tested by 100 test PSF--Cubes.}
   \label{figure008}
   \end{figure}

\begin{figure}
   \centering
   \includegraphics[width=\hsize]{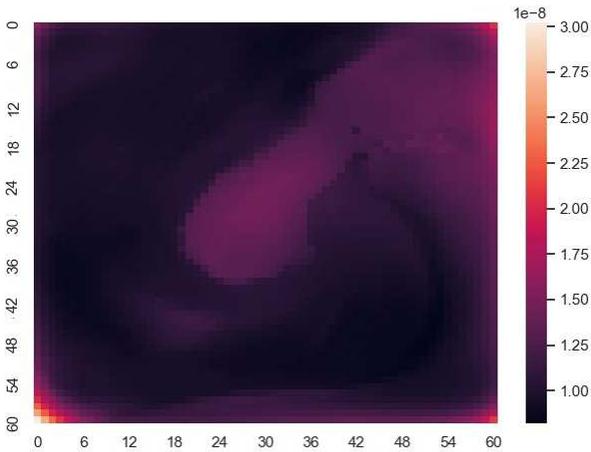}
      \caption{The heat map of mean MSE between original PSFs and PSFs reconstructed by the Tel--Net.}
   \label{figure009}
   \end{figure}

\begin{figure*}
\centering
\subfigure[Original PSF]{
\begin{minipage}[t]{0.33\linewidth}
\centering
\includegraphics[width=1\textwidth]{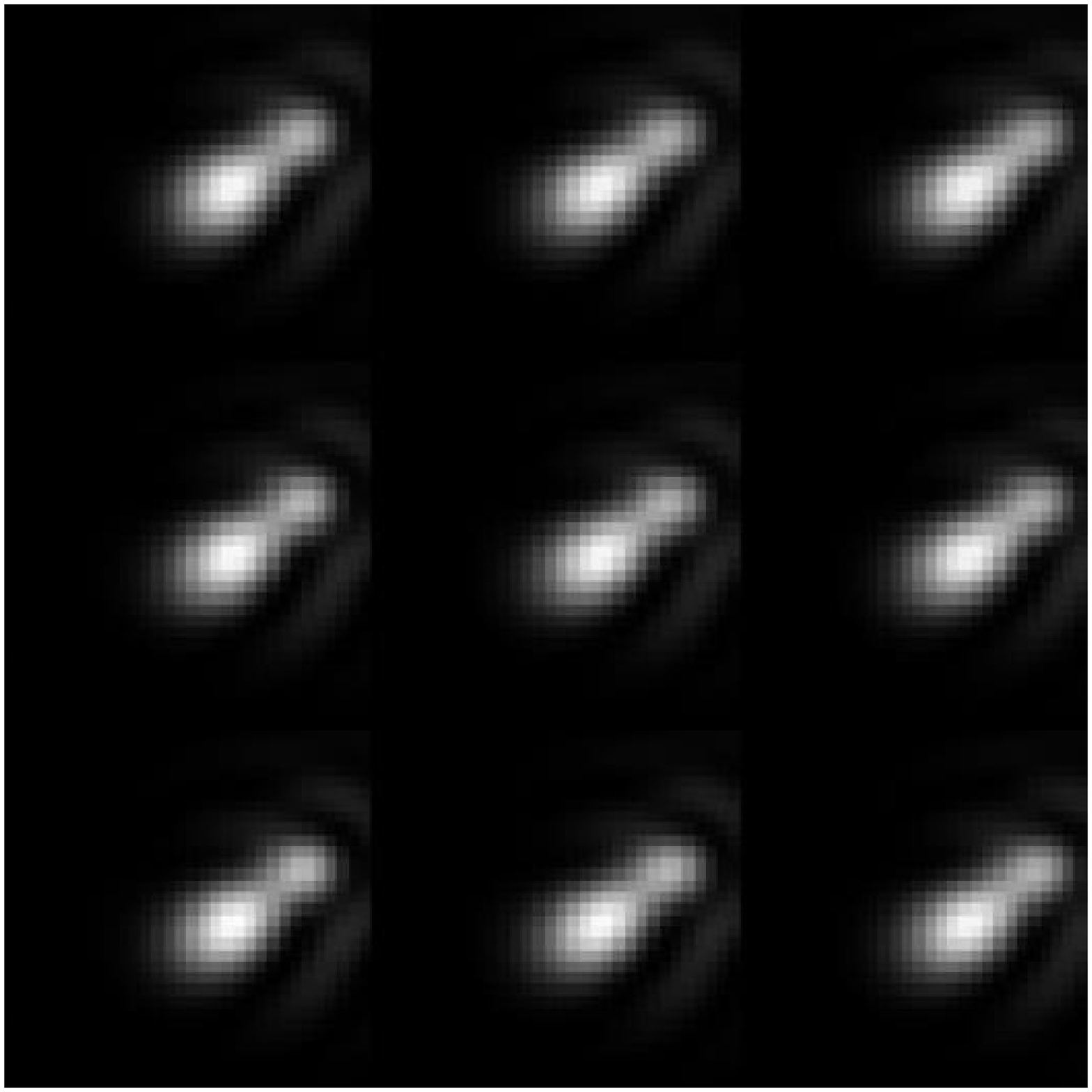}
\end{minipage}%
}%
\subfigure[Tel-Net predicted PSF]{
\begin{minipage}[t]{0.33\linewidth}
\centering
\includegraphics[width=1\textwidth]{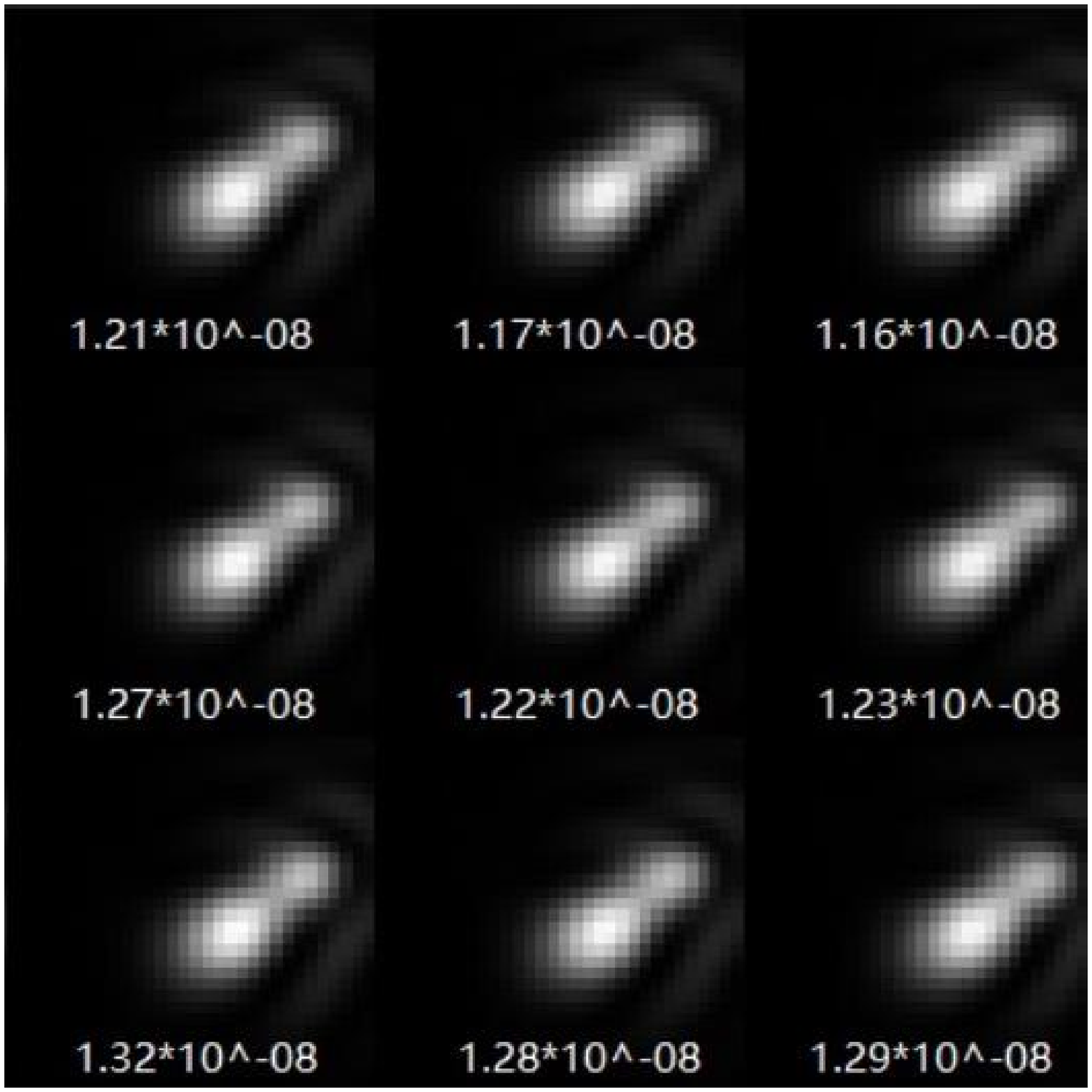}
\end{minipage}%
}%
\subfigure[IDW predicted PSF]{
\begin{minipage}[t]{0.33\linewidth}
\centering
\includegraphics[width=1\textwidth]{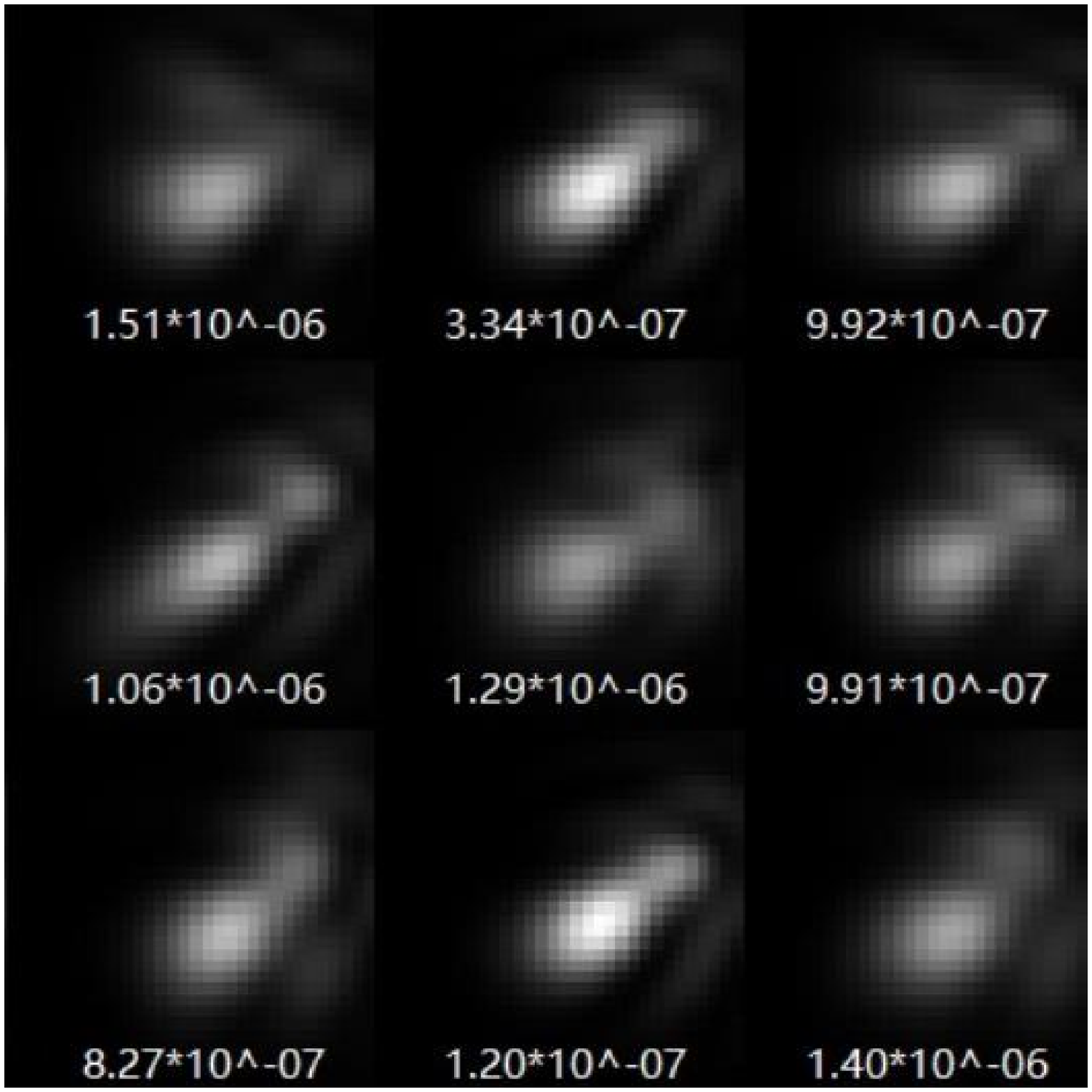}
\end{minipage}
}%
\centering
\caption{ Several examples of original PSFs and corresponding PSFs reconstructed by different methods with their MSE values, when the number of PSFs in the training set is large. (a) are original PSFs, (b) are PSFs predicted by the Tel--Net and (c) are PSFs predicted by the IDW with with four nearby PSFs.}
\label{figure010}
\end{figure*}

In the next step, we test the performance of the Tel--Net with a bigger dataset. This time we use the entire 2000 PSF-Cubes that are defined in the beginning of this section. We use 100 PSF--Cubes with random misalignments of the secondary mirror as the test set. The Tel--Net is trained through 10 epochs which is enough to achieve reasonable abilities. The MSE of the training set and the test set of different epochs is shown in figure \ref{figure007}. The MSE mean values are also calculted for every 4 epochs. As shown in this figure, the Tel--Net converges after 40 epochs. The statistical results of the MSE are shown in table \ref{table:PSFMSE2} and the histograms of the MSE are shown in figure \ref{figure008}. The results show that the Tel--Net has better ability to model PSFs in fine details than the interpolation method. The residual error is more than one order smaller than the IDW method. It indicates us that with adequate sampled PSFs as training set, the Tel--Net could almost predict any PSFs of WFSATs in any states. In comparison to results obtained on the smaller data set, we can find that the mean MSE has been reduced by half. The heat map of MSE in different positions of the FoV is shown in figure \ref{figure009}. We can find that with more PSFs as training set, the Tel--Net can better predict unknown PSFs. We show original and reconstructed PSFs in figure \ref{figure010}. As shown in this figure, we can find that with more PSFs as training set, the Tel--Net could better reconstruct PSFs.\\


	\begin{table}
	\caption{MSE between predicted PSFs (estimated by the Tel--Net and the IDW) and original PSFs. The Tel--Net is trained by 2000 training PSF--Cubes and tested by 100 test PSF--Cubes.}                 
	\centering          
	\begin{tabular}{c c c c}     
	\hline\hline                         
	 Method &MSE mean&MSE variance \\
	\hline   
	Tel-Net &$1.49  \times 10^{-8}$&$5.81  \times 10^{-17}$\\
	IDW &$7.79  \times 10^{-7}$&$2.04  \times 10^{-13}$\\
	\hline           
	\end{tabular}
	\label{table:PSFMSE2}
	\end{table}

\subsection{Training and testing the Tel-Net with PSFs of different size}
\label{sec:simresultdiffp}
To cover a large FoV in a cost effective way, WFSATs used in fast sky survey would obtain images with low spatial sampling rate. In these images, PSFs have very limited number of pixels. Meanwhile, for photometry sky surveys, PSFs of WFSATs may be well sampled (normally each PSF has tens to hundreds pixels). To test robustness of the Tel--Net in processing images with different number of pixels, we generate PSFs of different sampling rates ($32\times 32$ pixels, $16 \times 16$ pixels and $8\times 8$ pixels). The number and the strategy we used in this subsection is the same as those used in subsection \ref{sec:simresultbig}. The results of MSE for reconstructed PSFs of different sampling rates is shown in table \ref{table:pixels}. We can find that PSFs with different sampling rates can be well reconstructed by a trained Tel--Net. Meanwhile, as the pixel number increases, the MSE would decrease.
	\begin{table}
	\caption{MSE between predicted PSFs (estimated by the Tel--Net) and original PSFs for PSFs with different sizes. The Tel--Net is trained by 1900 training PSF--Cubes and tested by other 100 test PSF--Cubes.}                 
	\centering          
	\begin{tabular}{c c c c}     
	\hline\hline                         
	Number of pixels (pixelscale) &MSE mean&MSE variance \\
	\hline   
	$32  \times 32$ (0.54 arcsec/pixel) &$1.49  \times 10^{-8}$&$5.81  \times 10^{-17}$\\
	$16  \times 16$ (1.08 arcsec/pixel) &$3.03  \times 10^{-8}$&$1.03  \times 10^{-16}$\\
	$8  \times 8$ (2.16 arcsec/pixel)&$8.52  \times 10^{-8}$&$5.72  \times 10^{-16}$\\
	\hline           
	\end{tabular}
	\label{table:pixels}
	\end{table}

\subsection{Training and testing the Tel-Net with star images of different SNR}
\label{sec:simresultdiffsnr}
Since the aperture size of WFSATs is small,  images obtained by WFSATs may have relatively low SNR. Stars with low SNR would introduce additional error to the Tel--Net. To test the robustness of Tel--Net, we add different levels of Poisson noise to simulated PSFs ($32 \times 32$ pixels) as observed star images. Then, these simulated star images are used as PSFs to train and test the Tel--Net. The results are shown in table \ref{table:noise}. Because PSFs are all normalized to 1 for the Tel--Net, we multiply 10 or 80 with PSFs to generate star images with different levels of signal and then add the same level of Poisson noise as background noise. Therefore, 80 stand for star images with photon counts that are 8 times larger than stars with 10. As shown in this table, we can find that as noise levels increases, the performance of the Tel--Net would drop down. But the performance is still better than the IDW method.
	\begin{table}
	\caption{MSE between predicted PSFs (estimated by the Tel--Net and the IDW) and original PSFs for stars with different SNRs.}                 
	\centering          
	\begin{tabular}{c c c c}     
	\hline\hline                         
	Counts of photons&Methods &MSE mean&MSE variance \\
	\hline   
	$80$&Tel-Net &$2.77  \times 10^{-8}$&$3.12  \times 10^{-16}$\\
	$80$&IDW&$7.04  \times 10^{-3}$&$3.84  \times 10^{-5}$\\
	$10$&Tel-Net &$4.18  \times 10^{-8}$&$7.69  \times 10^{-16}$\\
	$10$&IDW&$9.62  \times 10^{-3}$&$4.32  \times 10^{-5}$\\
	\hline           
	\end{tabular}
	\label{table:noise}
	\end{table}
   
\section{Training and testing Tel--Net with experimental data}
\label{sec:realresult}
\begin{figure}
	\includegraphics[width=\columnwidth]{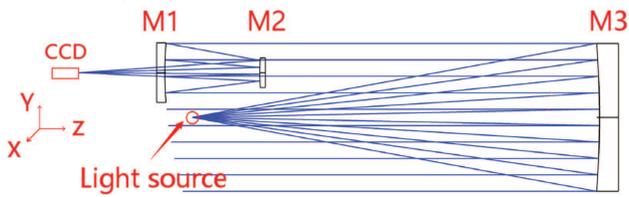}
    \caption{The concept design of the optical path in our experiment.}
    \label{figure011}
\end{figure}

\begin{figure}
	\includegraphics[width= 0.7\columnwidth]{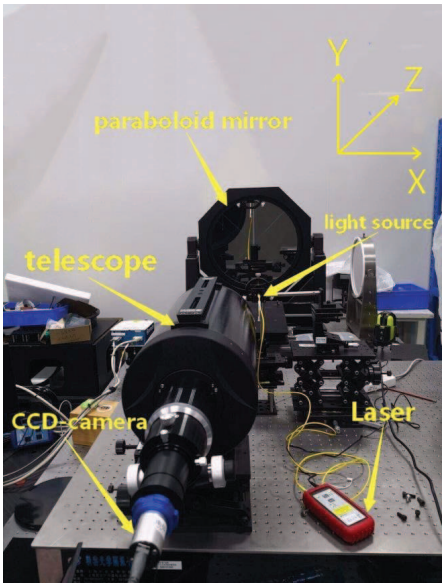}
    \caption{The optical path utilized in our experiment.}
    \label{figure012}
\end{figure}

\begin{figure}
   \centering
   \includegraphics[width=\hsize]{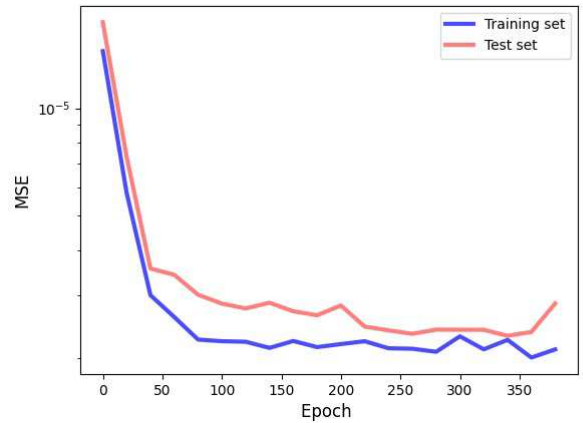}
      \caption{MSE of PSFs in the training set and the test set as function of epochs for real observation PSFs. As we can find from this figure, the Tel--Net is over--fitting, because there is only very limited number of PSF-Cubes.}
   \label{figure013}
   \end{figure}

To further test the performance of the Tel--Net, we set up an optical experimental bench in the laboratory according to the schema given in figure \ref{figure011} and with the experiment setup presented in figure \ref{figure012}. It includes a parabolic mirror (M3), an Ritcher-Chretien (RC) telescope with wavefront error (RMS value) less than $1/30 \lambda$ (M1 and M2), a fibre laser as a point-like source of light and a camera (CCD). We tilt the parabolic mirror (M3) to introduce additional aberrations and then move the camera together with the telescope to obtain PSFs in different positions of the FoV. The experiment is carried out in the optical laboratory located in Nanjing Institute of Astronomical Optics \& Technology, Chinese Academy of Sciences. In the laboratory, the temperature variation is smaller than 5 degrees and and the variation of humidity is below $10\%$. It took around 6 days to obtain all PSFs and in this paper, we assume variations of temperature and humidity would not introduce signifiant variations of PSFs. In total we generate 40 different levels of M3 tilts. For each levels of tilt, we obtain 25 PSFs equally distributed in $5\times 5$ patches. An exemplary PSF with aberrations obtained at the edge of the FoV is shown in figure \ref{figure013}.\\

Because automatic alignment devices such as electric precise lifting platforms are not used in this experiment, tilt of M3 and capture of PSFs are all done manually. Therefore, although we have spent a long time to collect PSFs, only very limited data are obtained, comparing with the amount of data generated previously by ZEMAX. Besides, due to limitations of human beings, these data have obvious imperfections: different levels of tilts are not strictly equally distributed and PSFs of the same patch corresponding to different tilts have position uncertainties up to several pixels. These problems impose the need for the use of automatic alignment instruments and related control methods while calibrating the smart data processing algorithms.\\

\begin{figure*}
\centering
\subfigure[Original PSF]{
\begin{minipage}[t]{0.6\columnwidth}
\centering
\includegraphics[width=0.6\columnwidth]{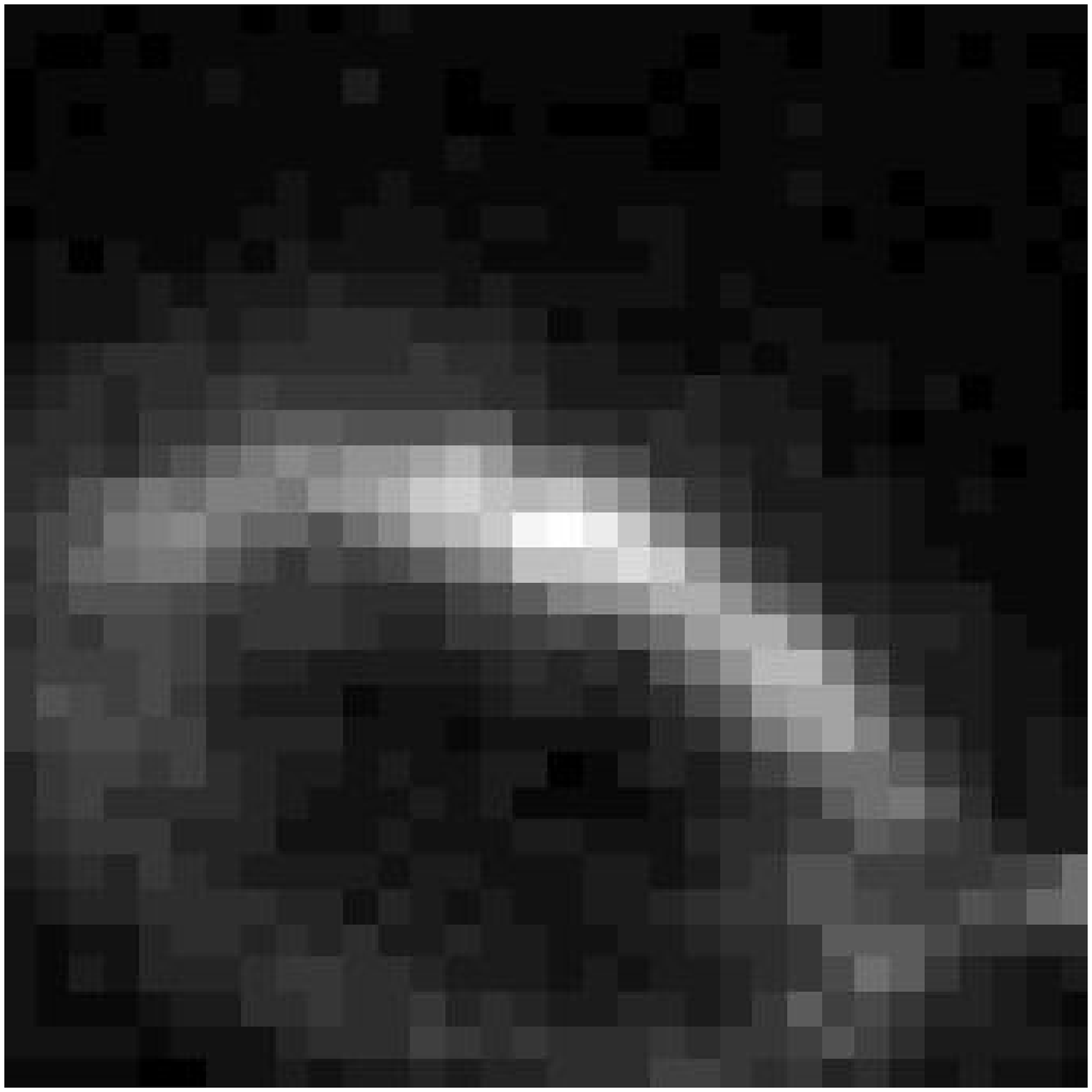}
\end{minipage}%
}%
\subfigure[Tel-Net predicted PSF]{
\begin{minipage}[t]{0.6\columnwidth}
\centering
\includegraphics[width=0.6\columnwidth]{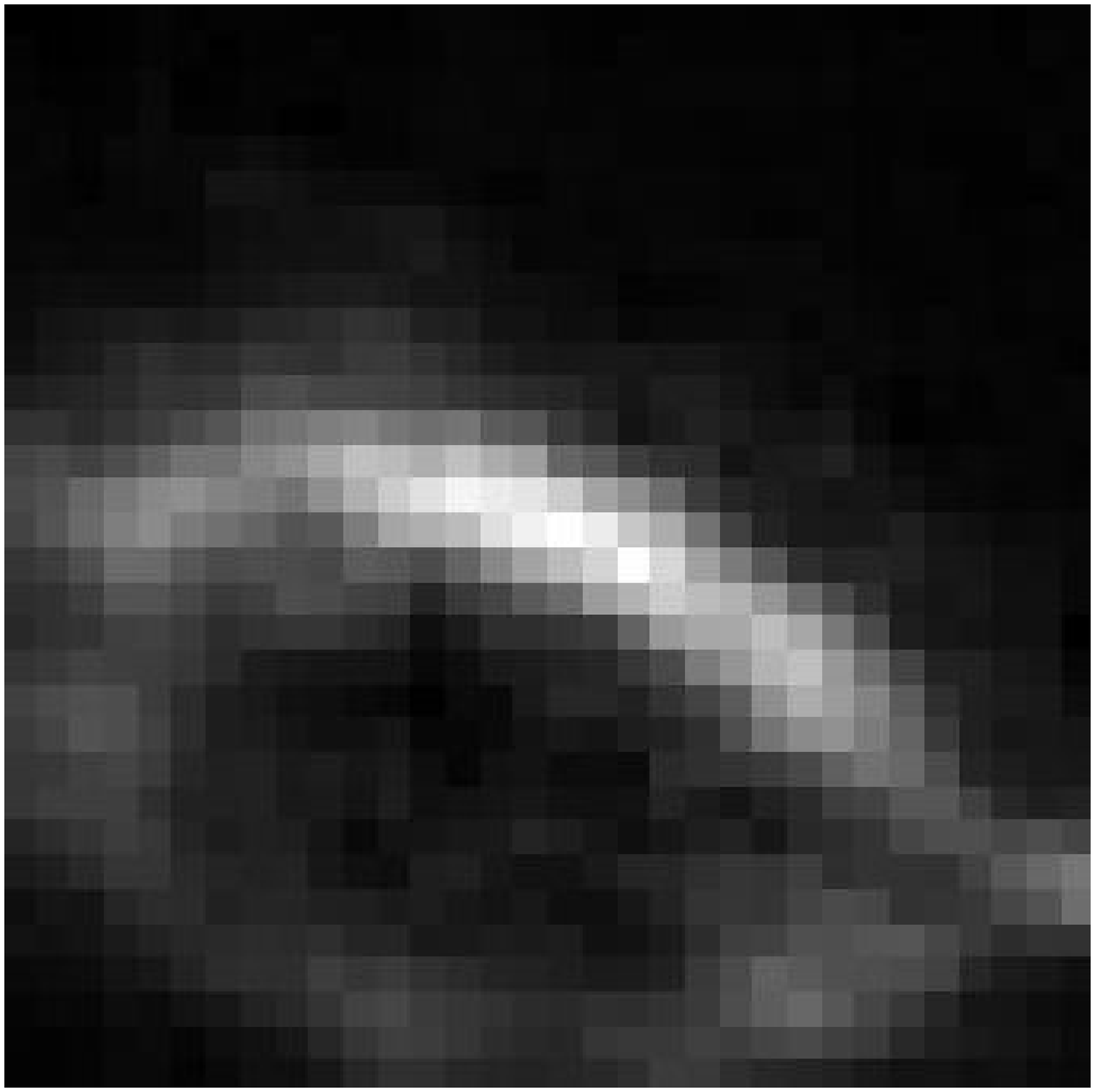}
\end{minipage}%
}%
\subfigure[IDW predicted PSF]{
\begin{minipage}[t]{0.6\columnwidth}
\centering
\includegraphics[width=0.6\columnwidth]{fig014c.eps}
\end{minipage}
}%
\centering
\caption{ (a) is the original PSF, (b) is PSF predicted by the Tel--Net and (c) is PSF predicted by the IDW with four PSFs as prior information.}
\label{figure014}
\end{figure*}

Besides, limited PSF data would introduce over--fitting problem. The relation between MSE and epochs is shown in figure \ref{figure013}. As shown in this figure, the Tel--Net converges after 250 epochs but we could find that the Tel--Net is over--fitting the training set. It is caused by limited number of PSF-Cubes, which would require automatic alignment devices to obtain more data in our future experiments. Although these PSF--Cubes have some issues, we are still able to train the Tel--Net with these data to test its robustness. In total, we obtain 40 PSF--Cubes and we randomly select 36 PSF--Cubes to train the Tel--Net. 8 PSFs are randomly selected as known PSFs in input PSF--Cubes. After training, we use the Tel--Net to predict PSFs with PSF--Cubes from the test set. One of PSFs predicted by the Tel--Net and that  predicted by the IDW method are shown in figure \ref{figure014}. The Tel--Net can predict PSFs in fine details, while the IDW method can only predict main structure of the original PSF. Kernel density estimation and statistical results of reconstruction error are shown in figure \ref{figure015} and table \ref{table3}. We can find that the Tel--Net has better performance than that of the IDW method, even with very limited data. It should be noted that because the number of PSFs in the training data is small, the results obtained in real applications are not as good as results obtained with the simulated data. This problem is caused by limited number of training data and could be solved through increasing the volumn of training data with applying automatic alignment devices when we obtain training data.\\

	\begin{table}
	\caption{MSE between predicted PSFs (estimated by the Tel--Net and the IDW) and original PSFs. The Tel--Net is trained by 36 training PSF--Cubes and tested by 4 PSF--Cubes.}                 
	\centering          
	\begin{tabular}{c c c c}     
	\hline\hline                         
	Method &MSE mean&MSE variance \\
	\hline   
	Tel-Net &$1.29  \times 10^{-6}$&$3.07  \times 10^{-12}$\\
	IDW &$2.83  \times 10^{-6}$&$5.60  \times 10^{-12}$\\
      \hline       
	\end{tabular}
	\label{table3}
	\end{table}
   \begin{figure}
   \centering
   \includegraphics[width=\hsize]{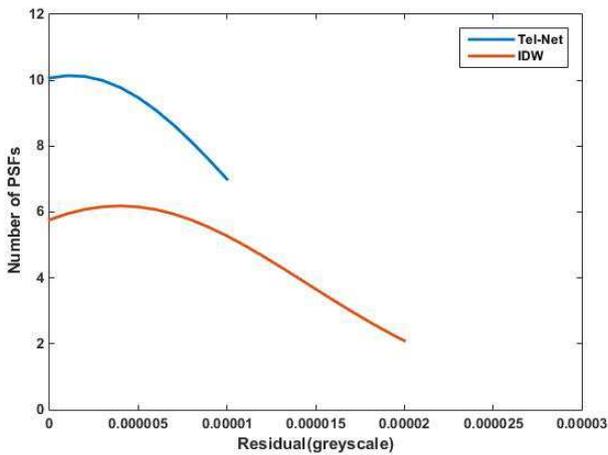}
      \caption{Histogram of MSE between PSFs predicted by Tel--Net and original PSFs. To make the results easier to compare, we use the kernel density estimation method to fit the Histogram. The blue curve stands for MSE of PSFs predicted by the Tel--Net and orange curve stands for MSE of PSFs predicted by the IDW method. The Tel--Net has better performance in PSF estimation.}
   \label{fig015}
   \end{figure}
\section{Conclusions and future work}
\label{sec:con}
In this paper, we propose the Tel-Net as a new PSF modelling framework for WFSATs. The Tel-Net assumes that PSFs in WFSATs can be understood as a response of a complex system for a single state of an imaging instrument. States of the whole optical system are sampled by PSF--Cubes during system test procedure. Then we train the Tel-Net with PSF--Cubes and use it to predict PSFs. We test the Tel-Net with simulated and real data and find that the Tel-Net can obtain PSFs in any positions of the FoV with finite number of star images. An example of classic PSF estimation method, the interpolation method utilizing the Inverse Distance Weighting has significantly worse performance and is not able to predict complex aberrations seen in both real and simulated PSFs.\\

The Tel-Net could be added as a necessary part to the calibration method for WFSATs, particularly for WFSATs that are lack of maintenance during observations, such as space telescopes. The Tel--Net indicates the importance of alignment data of a WFSAT pointing out the need for automatic characterization of the optical instrument during commissioning phase. In the future, we will use the Tel--Net as PSF modelling method to modify astronomical target detection and photometry neural networks \citep{Jia2020} to increase the observation accuracy of these telescopes when there are PSF variations in data obtained by these WFSATs.\\

\section*{Acknowledgements}
Authors would like to thank the reviewer, Dr. Morgan, Robert for his kindly suggestions that improve this paper a lot. Peng Jia would like to thank Dr. Alastair Basden from Durham University, Professor Rongyu Sun from Purple Mountain Observatory who provide very helpful suggestions for this paper. This work is supported by National Natural Science Foundation of China (NSFC)(11503018), the Joint Research Fund in Astronomy (U1631133, U1931207) under cooperative agreement between the NSFC and Chinese Academy of Sciences (CAS). Adam Popowicz was supported by grants: 02/140/RGJ21/0012 and BK-225/RAu-11/2021. Authors acknowledge the French National Research Agency (ANR) to support this work through the ANR APPLY (grant ANR-19-CE31-0011) coordinated by B. Neichel. This work is also supported by Shanxi Province Science Foundation for Youths (201901D211081), Research and Development Program of Shanxi (201903D121161), Research Project Supported by Shanxi Scholarship Council of China, the Scientific and Technological Innovation Programs of Higher Education Institutions in Shanxi (2019L0225). \\
Data Availability Statements: the code in this paper can be downloaded from aojp.lamost.org and after acceptance the code will be released in PaperData Repository powered by China-VO with a DOI number.\\




\bibliographystyle{mnras}
\bibliography{DAE} 








\bsp	
\label{lastpage}
\end{document}